\documentclass[%
 reprint,
 superscriptaddress,
nofootinbib,
 amsmath,amssymb,
 aps,
]{revtex4-2}

\usepackage{graphicx}
\usepackage{dcolumn}
\usepackage{bm}
\usepackage{color}

\usepackage[pdfpagelabels, pdfencoding=auto, psdextra]{hyperref}
\usepackage{hyperref}
\hypersetup{%
 pdfsubject=Paper,
 pdfkeywords={neutron stars, asteroseismology},
 unicode = true,
 breaklinks = true,
 colorlinks = true,
 linkcolor = blue,
 menucolor = blue,
 citecolor = blue,
 urlcolor = blue
}

\begin{document}

\preprint{APS/123-QED}

\title{Measuring radii of merging neutron stars with interface-mode asteroseismology informed by nuclear theory and experiment}

\author{Duncan Neill}
 \email{Contact author: dn431@bath.ac.uk}
\affiliation{
Department of Physics, University of Bath; Claverton Down, Bath BA2 7AY, UK. 
}

\author{William G. Newton}
\affiliation{
Department of Physics and Astronomy, East Texas A\&M University; Commerce, Texas 75429-3011, USA.
}

\author{Jeremy W. Holt}
\affiliation{
Cyclotron Institute, Texas A\&M University; College Station, Texas 77843, USA.
}
\affiliation{
Department of Physics and Astronomy, Texas A\&M University; College Station, Texas 77843, USA.
}

\author{Christian Drischler}
\affiliation{
Department of Physics and Astronomy, Ohio University; Athens, Ohio 45701, USA.
}
\affiliation{
Facility for Rare Isotope Beams, Michigan State University; East Lansing, Michigan 48824, USA.
}

\author{J\'er\^ome Margueron}
\affiliation{
International Research Laboratory on Nuclear Physics and Astrophysics, Michigan State University and CNRS; East Lansing, MI 48824, USA.\\
}

\author{David Tsang}
 \email{Contact author: d.tsang@bath.ac.uk}
\affiliation{
Department of Physics, University of Bath; Claverton Down, Bath BA2 7AY, UK. 
}

\date{\today}

\begin{abstract}
The structure and dynamics of neutron stars can be used to probe the physics of extreme matter at nuclear densities and beyond. Nucleonic matter up to $\sim2-3$ times nuclear saturation density is well-studied by nuclear experiments and theoretical modelling. Matter beyond these densities may contain non-nucleonic degrees of freedom that determine the structure of the neutron star inner core and influence bulk observables like stellar radius. Neutron star radius is a key parameter for constraining the core equation of state, but is not a direct gravitational-wave observable during neutron star mergers. Here we show that, if nucleonic physics is well constrained at low densities, the frequency of the asteroseismic crust-core interface mode in a neutron star can be used to infer its radius to within 5-10\%, in a way which is notably insensitive to the details of the inner core. This frequency can be measured through multimessenger coincident timing of resonant shattering flares, or direct observation of dynamical tidal resonance with next-generation gravitational-wave detectors. We show that improved constraints on low-density nucleonic physics by nuclear experimental and theoretical efforts will substantially improve such a radius measurement, leveraging low-density efforts for an improved understanding of physics at higher densities.

\end{abstract}

\maketitle

\section{Introduction}

Neutron stars are compact stellar remnants where cold neutron-rich matter surpasses nuclear densities on macroscopic scales. Observations that inform us of neutron star structure and dynamics therefore provide unique insight into the physics of dense nuclear matter, probing conditions that can not yet be produced in terrestrial laboratories. However, uncertainty in the phases and structures of matter present in neutron stars may add significant systematic uncertainty to inferences of dense matter from astronomical observables. Understanding the scope of these uncertainties and identifying observables that are independent of them is important for obtaining reliable constraints on dense matter.

\subsection{Neutron Star Matter from Low to High Densities} 
The structure of neutron star (NS) matter at baryon densities below $\sim$2 times nuclear saturation density ($n_s\approx0.16 \text{ fm}^{-3}$) can be robustly predicted by theory. Matter there is nucleonic, with a solid crust extending down to a depth of $\sim1$ km composed of layers of nuclear clusters in a Coulomb lattice surrounded by electrons and, in the inner crust where the proton fraction is below neutron-drip, free neutrons. Around $\sim 0.5 n_s$, matter transitions to a fluid of nucleons, protons, electrons and muons: the outer core. Uncertainties in inner crust and outer core microphysics are primarily quantitative and reducible by nuclear modelling and experiment (see \citet{kumar_theoretical_2024} and references therein), including future neutron skin measurements at MESA \cite{becker_p2_2018} and Heavy-Ion Collisions at facilities such as FRIB \cite{balantekin_nuclear_2014} and FAIR \cite{durante_all_2019}. 

At densities above $2-3 n_s$ (the NS inner core), physics is qualitatively uncertain \cite{haensel_neutron_2007}; matter possibly consists of an admixture of hyperons and nucleons, meson condensates, quarkyonic matter, or deconfined quarks, with phase transitions of unknown nature between them. Matter at high density, low temperature, and high isospin asymmetry is inaccessible to terrestrial experiments and theoretical studies with controlled uncertainties. These conditions are, however, found in the inner cores of NSs, so astrophysical observables probing structure and dynamics of NSs can be used to constrain the properties of this extreme matter \cite{lattimer_neutron_2021,chatziioannou_neutron_2025}. 

\citet{lindblom_determining_1992} demonstrated that the NS mass-radius ($M_{\rm NS}-R_{\rm NS}$) relationship can be mapped 1-1 to the NS equation of state (EOS), providing a means by which precise radius measurements of isolated and accreting NSs \cite{riley_nicer_2019,riley_nicer_2021,choudhury_nicer_2024,salmi_nicer_2024,mauviard_nicer_2025} at different masses can constrain the NS EOS \cite{legred_impact_2021,huang_constraining_2025} and subsequently dense matter. An observable complementary to $M_{\rm NS}-R_{\rm NS}$ constraints is the tidal deformability ($\Lambda$): a measure of the response of NSs to tidal gravitational fields, which can be inferred from gravitational-wave (GW) observations of merging NS binaries. While the radius and tidal deformability are both strongly dependent on the EOS in the core regions of a NS, the $M_{\rm NS}-\Lambda$ to EOS mapping is known to be degenerate \cite{raithel_degeneracy_2023,raithel_tidal_2023}, particularly when core EOSs are allowed to contain phase transitions. Thus, radius constraints inferred from tidal deformability measurements \cite{de_tidal_2018} are dependent on core EOS assumptions and are not robust to the core's qualitative uncertainties \cite{legred_assessing_2024} or inconsistencies in how the core is modelled relative to the crust \cite{suleiman_influence_2021}. Radius measurements independent of tidal deformability aid in breaking the $M_{\rm NS}-\Lambda$ to EOS degeneracy, allowing for the EOS to be constrained over a wider range of densities than with either measurement alone.

In this work, we will show how a measurement of the NS quadrupolar crust-core interface asteroseismic mode's frequency in a binary merger would allow us to infer NS radius independent of core uncertainties and tidal deformability, opening up the possibility of obtaining simultaneous measurements of $R_{\rm NS}$ and $\Lambda$ for the same star and thus reliable constraints on inner-core physics.

\subsection{Asteroseismology of neutron stars in the LIGO-Virgo-KAGRA era and beyond.} 

Tidal forces in NS-NS and black hole (BH)-NS mergers excite quadrupolar oscillation modes in the NS(s), which extract energy from the binary orbit through off-resonant and resonant dynamical tides \cite{lai_resonant_1994} and thus produce phase shifts in gravitational-wave (GW) signals from these mergers.
Current LIGO-Virgo-KAGRA (LVK) detectors lack sufficient sensitivity for robust direct dynamical tide detection \cite{andersson_using_2018,pratten_gravitational-wave_2020}, but may still indirectly offer insight into mode properties through multimessenger observations. Resonant shattering flares (RSFs) are brief gamma-ray flares triggered by the NS crust fracturing, which may occur when an oscillation mode is resonantly excited by the NS-NS or BH-NS orbit and exceeds the NS crust's elastic limit \cite{tsang_resonant_2012,neill_resonant_2021,most_nonlinear_2024}. The GW chirp frequency at the moment of RSF emission directly corresponds to the resonant mode's frequency, so the coincident detection of an RSF by instruments such as Fermi/GBM and GWs by the LVK network would provide an asteroseismic constraint on the frequency of the resonant mode. 

Looking beyond current facilities, next-generation GW facilities -- Einstein Telescope (ET) and Cosmic Explorer (CE) -- will greatly enhance our ability to perform NS asteroseismology. The improved sensitivities of these facilities to GW signals from binary mergers will allow us to directly resolve asteroseismic spectra displaying multiple NS modes without coincident electromagnetic detection, while also observing such signals more frequently than at present. Different oscillation modes are sensitive to a variety of physics and features of NS structure, so constraints on their frequencies and eigenfunctions provide a variety of detailed complementary probes of NS interior physics. Certain modes are concentrated within small density ranges, allowing facilities like ET and CE to study properties of matter and NS structure in more localized regions of the star \cite{counsell_interface_2025}. Efforts are underway to understand the effects of mode excitations on GWs and develop waveform models describing them \cite{read_waveform_2023,hinderer_effects_2016,steinhoff_dynamical_2016,schmidt_frequency_2019}.

The quadrupolar crust-core interface-mode (i-mode) is a strong candidate for triggering RSFs \cite{tsang_resonant_2012} and may be sufficiently excited during merger to be detected with next-generation GW facilities \cite{passamonti_dynamical_2021}. 
This mode arises in NSs due to the phase transition at the crust-core boundary and exists as a trapped oscillation that propagates around the star at this transition, and is therefore strongly influenced by properties of matter there. Previous studies have demonstrated its sensitivity to local properties such as buoyancy and elasticity \cite{tsang_resonant_2012,neill_resonant_2021,passamonti_dynamical_2021,mcdermott_nonradial_1988,gittins_neutron-star_2025}, but in this work we will examine its dependence on global properties of the NS at the crust-core transition: the radius of the transition (which determines the geometric size of the trapping region), and the enclosed mass (which with radius determines the local gravitational acceleration). We will begin by describing our NS modelling and how it allows the NS inner core to vary independently of physics at lower densities, and then show that the relationship between i-mode frequency and NS radius is independent of the inner-core model. We will then add realistic uncertainties to the physics at lower densities to determine how well this relationship would allow radius to be constrained by a measurement of i-mode frequency.

\section{NS model construction}

The following is a brief overview of the methods we use to construct NS models throughout this work. For further detail, see supplementary material.

We construct spherically symmetric equilibrium NS models following the NS meta-modelling approach of \citet{margueron_equation_2018}, \citet{grams_properties_2022} and \citet{margueron_equation_2018-1}, neglecting effects of rotation, magnetic fields and finite temperatures. The crust is constructed using the Compressible Liquid Drop Model (CLDM), wherein we assume matter to be composed of identical unit cells treated as spherical under the Wigner-Seitz approximation. Each cell contains a nuclear cluster surrounded by pure neutron matter in chemical and mechanical equilibrium, both of which are uniform in density, separated by a surface whose energy must be specified. The crust-core transition density is where the energy per baryon of crust and uniform matter cross over. The properties of clusters, free neutrons and uniform core matter are then determined by the dense nuclear matter EOS: energy per baryon as a function of nuclear asymmetry and density. 
By expanding this EOS around symmetric nuclear matter (SNM) we obtain the EOS of SNM and the symmetry energy ($E_{\rm sym}$, the difference between pure neutron matter (PNM) and SNM), each of which we further expand around $n_s$ as Taylor series to obtain a set of isoscalar (SNM) and isovector ($E_{\rm sym}$) parameters describing the EOS. These parameters play a key role in determining the equilibrium structure of the NS.

To explore the effect of non-nucleonic degrees of freedom in the NS inner-core, for densities above a transition at $n_t\geq1.5n_s$ we replace our NS EOSs with a range of parametric models that allow the NS EOS to vary widely, independent of the nucleonic physics at lower densities.
In addition to varying the parameters of the core model over wide ranges, we use several different types of model to determine whether our results are dependent on the method used to construct the core: models with first-order phase transitions (FOPT) followed by linear sound speed squared, 2- or 3-segment piecewise polytropic models, 3-segments linear sound speed squared models, and models featuring continuous sound speed that peaks and then approaches the high-density limit of perturbative QCD from below \cite{tews_constraining_2018}. Alongside the EOS, we must also choose descriptions for compositional properties of the inner core that affect oscillations: buoyancy and elasticity. Buoyancy driven by matter stratification is effectively suppressed by equilibrating processes (e.g., beta decay), which damp oscillations \cite{reisenegger_new_1992}. These processes are slow in cold nucleonic NS matter (relative to the oscillations of interest to this work), but in exotic phases of matter they may be significantly faster \cite{baym_neutron_1995}, so we set compositional buoyancy to zero throughout the inner core. For elasticity meanwhile, we assume that the core is a fluid and thus does not support shear forces, setting the shear modulus to zero.

\section{Dependence of i-mode frequency on the NS inner core}

\begin{figure}
\includegraphics[width=\columnwidth]{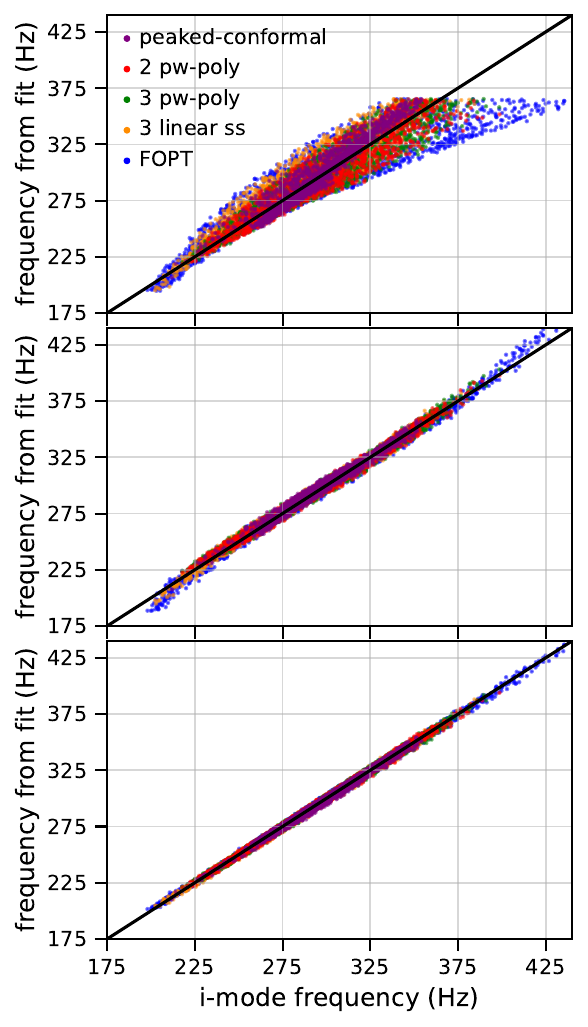}
\caption{\label{fig:frequency_fit} 
Results of i-mode frequency fits using bulk NS properties for NS models built from the same low-density nucleonic matter model, only allowing the NS inner core to vary. Note that with a fixed crust model, the total mass and radius of the NS are one-to-one with the core mass and crust-core transition radius. The fitting formulae are functions of NS mass (top), NS mass and radius (middle), and NS mass, radius, and a weighted average of compositional buoyancy in the core (bottom) (see Appendix~\ref{sec:appendix_fits}), and are fit to an equal number of samples from each of several different core model constructions (see text and legend). The fits work well for all of the core constructions we use, indicating that the relationship between i-mode frequency and these bulk NS properties is robust. We see that with the example fixed low-density model used here, i-mode frequency is well fit while only considering NS radius and mass, meaning that details of the inner core are not required. Including buoyancy marginally improves these fits, although it can be more significant for other fixed low-density nucleonic physics models with higher degrees of stratification (and thus greater compositional buoyancy) in the outer core (see supplementary material for examples).}
\end{figure}

To isolate the i-mode's dependence on the NS inner core, we fix the isoscalar and isovector nuclear matter parameters to fiducial values to obtain NS models with identical nucleonic components at low densities but with a wide range of inner cores. For each NS model, we sample a NS mass from 0.6 to 3.0 solar masses ($M_{\odot}$) and calculate the i-mode frequency assuming the relativistic Cowling approximation \cite{yoshida_nonradial_2002}. We then identify optimal fitting formulae for the i-mode frequency (see Appendix~\ref{sec:appendix_fits}) that progressively incorporate NS mass, NS radius, and a weighted average of the (compositional) Brunt frequency (the frequency of oscillations due to buoyancy) in the NS core. The resulting fits are shown in Fig.~\ref{fig:frequency_fit}. Fitting with mass alone provides a rough estimate for i-mode frequency, but adding radius information 
results in a highly predictive fit,
indicating that the i-mode frequency's core-dependence is mostly described by just the bulk NS mass and radius, regardless of details of the inner core. 
We see that the fits are of similar quality for each of the core models we have used, so this is not simply a consequence of model construction.

Having shown that the i-mode frequency is related to NS radius and is insensitive to details of the NS core model, we perform Bayesian inferences to examine the radius constraints that can be obtained from synthetic i-mode frequency measurements. For simplicity, we only use our first order phase transition (FOPT) core model construction, which gives the widest range for NS mass and radius of any of our core model types. Rather than defining the prior parameter distribution entirely in this model's parameter space, which non-trivially determines the shape of the prior in NS mass and radius and thus will affect the radii we recover \cite{sun_new_2026}, we choose a prior that is uniform in mass and radius. This is done by preliminarily sampling the model parameters uniformly within broad bounds (see supplementary material) and NS mass over a 0.6-3.0 $M_{\odot}$ range to obtain the distribution of radii given by uniform sampling; our prior is then obtained by weighting uniform samples inversely to a kernel density estimate of this mass-radius distribution.

In Fig.~\ref{fig:inferences} (left column), we demonstrate the steps of constructing NS models and inferring the radius. The top panel illustrates the fixed EOS of dense nuclear matter used in this example (split into SNM and PNM). The 2nd panel shows a sample of mass-radius curves obtained when various FOPT inner-core models are attached to the resulting low-density nucleonic NS model, and the third panel shows the i-mode frequencies and NS radii for a corresponding sample of NSs with masses in the range 0.6-3.0 $M_{\odot}$. In the 2nd and 3rd panels, an example of the mass range and samples to which the lower-mass NS\footnote{Following the argument of \citet{neill_resonant_2022}, for NS-NS mergers we assume that the mode resonance responsible for an RSF occurs in the lower-mass NS, and thus that the lower-mass NS is more likely to have its i-mode frequency measured.} in a NS-NS merger could be constrained by GWs is highlighted; to ensure a realistic range, we use the (low-spin) GW170817 mass posteriors \cite{abbott_GW170817_2017}. Finally, the bottom panel shows our NS radius distribution given the GW170817 mass constraint (labelled as our prior) and the posteriors recovered by adding synthetic i-mode frequency data. For an injected frequency normal distribution with standard deviation of 1\% (3\%) [5\%], we recover a radius posterior with 1$\sigma$ uncertainty of 0.277 km (0.500 km) [0.735 km].

\section{Adding uncertainty in nucleonic physics}
Assuming perfect knowledge of matter outside the NS inner core presents an extremely optimistic scenario for radius recovery; the i-mode is most sensitive to local properties of matter near the crust-core transition \cite{neill_resonant_2021}, so even modest uncertainties at low densities could mask the influence of NS radius on i-mode frequency. To examine our ability to recover NS radii when realistic uncertainties are included, we construct two samples of nucleonic meta-models: those with isoscalar and isovector parameter ranges informed by predictions from nuclear theory, and those informed by measurements from nuclear experiment. We also explore how future advances in our understanding of dense matter will improve the methods presented in this work by artificially reducing these uncertainties. 

\begin{figure*}
\includegraphics[width=\textwidth]{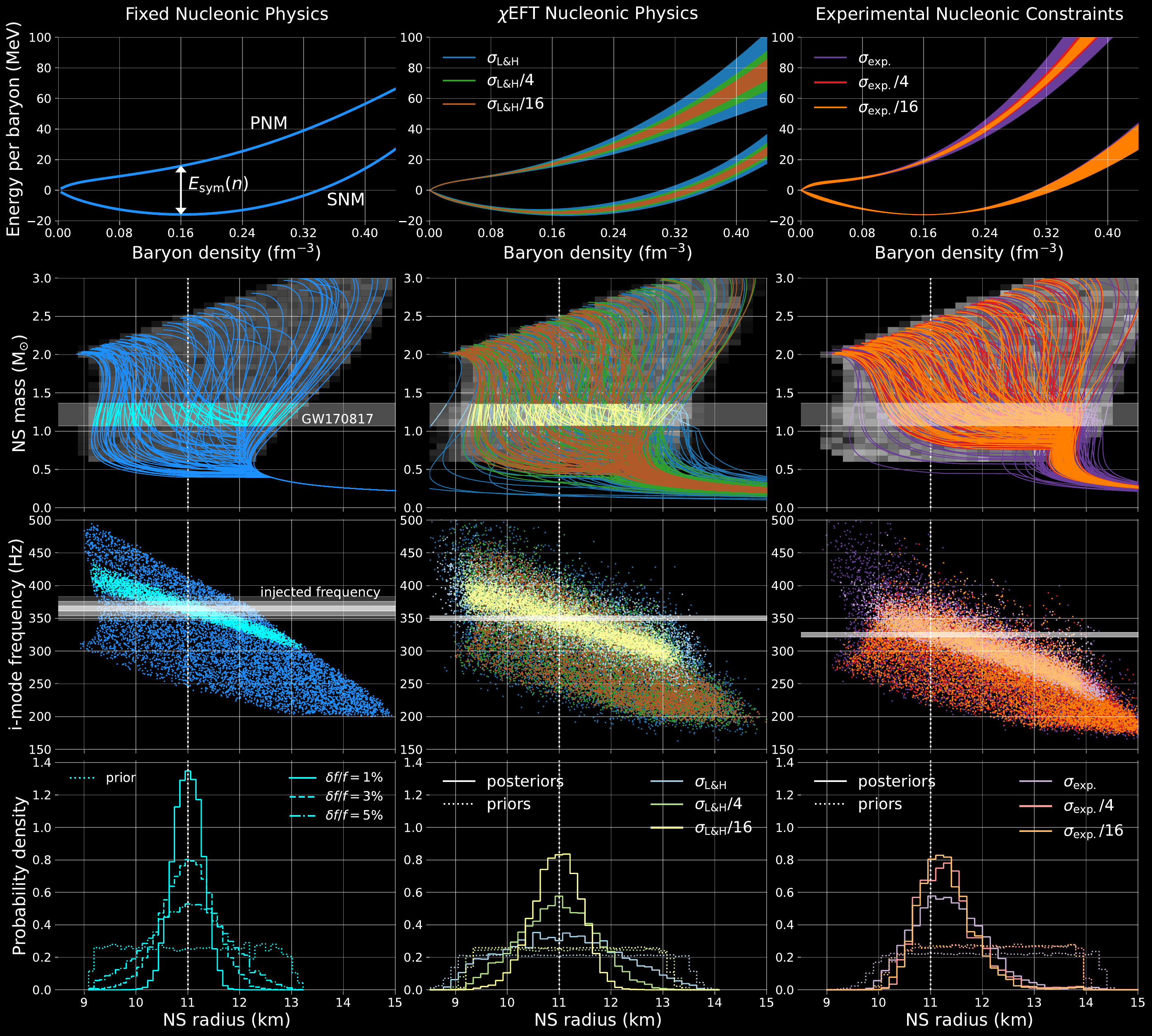}
\caption{\label{fig:inferences} Inference of NS radius given different nucleonic physics constraints and i-mode frequency measurements. \textbf{Left Column:} Fixed low-density nucleonic matter; \textbf{Middle Column:} Theoretical $\chi$EFT constraints on nucleonic matter (blue) from \citet{lim_neutron_2018}, with uncertainties scaled by factors of 1/4 (green), and 1/16 (brown/yellow); \textbf{Right Column:} Current experimental constraints on nucleonic matter (purple), based on \citet{tsang_determination_2024}, \cite{oliinychenko_sensitivity_2023} and \cite{schwenk_resonant_2005}. Additional constraints are added that reduce the uncertainties in the symmetry energy at 0.12 and 0.32 fm$^{-3}$ by factors of 1/4 (red), and 1/16 (orange). \textbf{Top row:} PNM and SNM energy per particle. Filled regions are between 1$\sigma$ bounds. \textbf{Second row:} Mass-radius curves for NS EOSs constructed using nuclear matter models from the upper panels and a wide range of parametric NS inner-core models. The representative mass uncertainty range for GW170817 is highlighted. The priors used when inferring NS radius are constructed such that samples are approximately uniform in radius at any given mass, shown in the background as a greyscale histogram. \textbf{Third row:} i-mode frequency vs NS radius, with darker colours showing a sample with a wide mass range (0.6-3 $M_{\odot}$) and lighter colours showing a sample with masses consistent with the low-spin posteriors of GW170817 \cite{abbott_GW170817_2017}. Also shown are 1$\sigma$ uncertainty regions for the injected i-mode frequency measurements used to generate the posteriors below (1\%, 3\% and 5\% for the left column, only 1\% for the middle and right). \textbf{Bottom row:} NS radius distributions. The injected model has a radius of 11 km (indicated with a white line). For fixed nucleonic physics, posteriors are shown for different frequency uncertainties, while for the theoretically- and experimentally-informed inferences we show the effect of improving current nuclear physics constraints. The dotted priors indicate results without any frequency measurement.}
\end{figure*}

Theoretical uncertainties:\; We draw samples from ab initio chiral effective field theory ($\chi$EFT) predictions with nucleon-nucleon and three-nucleon forces \cite{drischler_chiral_2021,epelbaum_high-precision_2020,machleidt_chiral_2011}. We use two sets of predictions – those from \citet{lim_neutron_2018} and those from \citet{drischler_quantifying_2020} – to explore systematic uncertainties, repeating our calculations for each separately. We find that the results are similar (see supplementary material) and so only show results for the former in Fig.~\ref{fig:inferences}.

Experimental uncertainties:\; We primarily follow \citet{tsang_determination_2024}, using their broad priors on the isoscalar and isovector parameters and informing our inferences by the asymmetric and symmetric matter constraints listed in their Table 1. The exceptions to this are that we use constraints based on the work of \citet{oliinychenko_sensitivity_2023} for the pressure of symmetric nuclear matter instead of those from \citet{tsang_determination_2024}, and that to avoid extremely soft PNM at low densities we add a constraint on it based on the calculations of \citet{schwenk_resonant_2005} for a unitary neutron gas.

Future improvements:\; For the theoretical constraints, we reduce the covariance matrices describing the uncertainty in the dense nuclear matter EOS by factors of 4 or 16. For the experimental constraints, we add new constraints that reduce uncertainty in the EOS at specific densities by factors of 4 or 16; we find that the best single such constraint is on the nuclear symmetry energy at $0.12 \text{ fm}^{-3}$, but also that simultaneously reducing the uncertainty in the symmetry energy at $0.12 \text{ fm}^{-3}$ and $0.32 \text{ fm}^{-3}$ results in much stronger radius constraints, so we use the latter. This choice is aligned with the projected experimental sensitivity of FRIB400, the proposed $400 \text{ MeV/u}$ energy upgrade of FRIB that is expected to provide complementary constraints on the symmetry energy in this density regime \cite{gade_frib400_2019}. Note that the improvements for theoretical and experimental constraints are qualitatively different, so their impacts should not be directly compared.

For more detailed descriptions of the sampling used, see Appendix~\ref{sec:appendix_NMEOS}. Similar to what we did for the fixed nucleonic model, we shape our priors to be uniform in mass and radius after sampling the isoscalar and isovector parameters and attaching a wide range of FOPT core models. This has little effect on the distribution of isoscalar and isovector parameters, as the parametric core of the NS affects radius more than the outer layers.

Fig.~\ref{fig:inferences} shows the stages of constructing models and inferring NS radius with nucleonic uncertainties from theory (middle column) and experiment (right column), where we have used synthetic frequency measurements with 1\% uncertainties that are peaked near the most probable value in the prior distribution for samples with radii of 11 km. Also shown are the results for the `improved' constraints. We see that when using current experimental constraints, while there is a larger spread in the relationship between i-mode frequency and NS radius than when the nuclear matter EOS was fixed (left column), radius information can still be extracted from measurements of the i-mode frequency, with the 90\% confidence interval decreasing from 4.07 km to 2.18 km (ending with a 1$\sigma$ uncertainty of 0.71 km). A smaller amount of radius information might be obtained when relying on theoretical predictions, with the interval going from 4.25 km to 3.50 km (ending with a 1$\sigma$ uncertainty of 1.07 km). Looking at what effects future improvements to the nuclear matter EOS constraints may have, we see that the radius 90\% confidence intervals tighten down to 1.62 km or 1.49 km (and 1$\sigma$ uncertainties down to 0.53 or 0.46 km) for 16 times stronger experimental or theoretical constraints (respectively), indicating that efforts in theory and experiment at low densities are crucial to maximising our ability to extract NS radii.

The uncertainties we have used for NS mass and i-mode frequency measurements are achievable with current facilities and may be common with 3rd-generation GW facilities. 
For the GW170817 posteriors we used for NS mass, the 1$\sigma$ uncertainty was approximately 5\% \cite{abbott_GW170817_2017}. With A+ facilities, we might expect $\sim0.2-2$ measurements to this precision or better per year \cite{the_ligo_scientific_collaboration_gwtc-40_2025}, while with multiple 3rd generation facilities, such precise measurements may be common and rare GW170817-like events may have mass uncertainty as low as 1\% \cite{puecher_measuring_2023}.
Conservative estimates place the frequency uncertainty from RSF timing at $\sim5-15\%$ \cite{neill_resonant_2021}, but detailed tidal energy-transfer analysis and direct GW detection with next-generation facilities will improve on this, perhaps even surpassing the $1-5\%$ range used in this work. 
For an overview of the relative impacts of uncertainties in NS mass, i-mode frequency and nucleonic physics on our radius recovery, see supplementary material.

\section{Conclusions}

We have found that measurements of the i-mode frequency through coincident RSF and GW timing, or directly from GW phase shifts with next-generation GW facilities, will provide strong constraints on NS radii. These constraints will only grow stronger as our understanding of dense nuclear matter at densities found outside the NS core improves, and with upcoming experiments \cite{becker_p2_2018,balantekin_nuclear_2014,durante_all_2019} and ongoing work to improve nuclear modelling, such improvements may be possible in the near future. NS radii constrain the EOS of the NS core, and so the method presented here provides an astrophysical framework where improvements in the study of low-density nuclear matter will enable more precise constraints at the high densities that cannot be directly probed in terrestrial nuclear experiments.

Comparing the radius posteriors we are able to obtain with synthetic data to radii inferred for isolated and accreting NSs \cite{riley_nicer_2019,riley_nicer_2021,choudhury_nicer_2024,salmi_nicer_2024,mauviard_nicer_2025} we see that asteroseismology will be competitive with the current state-of-the-art while providing access to the radius for a whole new population of NSs in binary mergers. Unlike constraints on radius inferred from tidal deformability, the method of extracting radius presented here is robustly independent of choices made in modelling the NS core. The lack of dependence on tidal deformability also allows for deeper insight into the nature of matter in the NS core, as radius and tidal deformability are complementary. No other robust radius measurements exist for merging NSs, exemplifying the new science unlocked by NS asteroseismology. This capability is currently limited and reliant on rare multimessenger events, but will be transformed by the increased event rates and direct resonant mode spectra made possible by the Einstein Telescope and Cosmic Explorer.

\begin{acknowledgments}
This work as initially arose from discussions during the `eXtreme Matter in eXtreme Stars (XMXS)’ workshop at the Lorentz Center (Leiden, The Netherlands), whom we gratefully acknowledge for their financial and organizational support of the workshop, which was also supported through grants from the (UK) Royal Astronomical Society, (French) Centre National de la Recherche Scientifique (CNRS), and Nederlandse Organisatie voor Wetenschappelijk Onderzoek (NWO) grant ENW-XL. We also thank the Institute for Nuclear Theory at the University of Washington for its kind hospitality and stimulating research environment during INT workshop 25-2b supported by the INT's U.S. Department of Energy grant No. DE-FG02- 00ER41132. D.N. and D.T. were supported by the UK Science and Technology Facilities Council (ST/X001067/1) and the Royal Society (RGS/R1/231499). W.G.N. was supported by National Science Foundation award PHY-2209536. The work of J.W.H. was supported in part by the US National Science Foundation under Grant No. PHY-2514930. C.D. acknowledges that this material is based upon work supported by the NSF under award PHY-2339043 and the U.S. Department of Energy, Office of Science, Office of Nuclear Physics, under the FRIB Theory Alliance award DE-SC0013617. J.M. acknowledges the support from CNRS-IN2P3 MAC2 masterproject and the project RELANSE ANR-23-CE31-0027-01 of the French National Research Agency (ANR). 
\end{acknowledgments}

\appendix

\section{Fitting i-mode frequency with bulk NS properties}\label{sec:appendix_fits}

\subsection{Mode calculation and i-mode identification}

We calculate the NS mode spectrum following the formalism of \citet{yoshida_nonradial_2002}, which is a general relativistic calculation for non-rotating spherical NSs in the relativistic Cowling approximation (i.e., neglecting perturbations of the space-time metric due to oscillating matter). We identify the quadrupolar crust-core i-mode as a single mode that has a cusp in radial displacement and a discontinuity in transverse displacement at the crust-core boundary, and that has no nodes in its radial displacement in the core. This mode is typically found in the $\sim100-600 \text{ Hz}$ range, between the g- and s-modes. 

In rare models where NS matter is highly stratified (and thus has unusually strong buoyancy), the frequency of the first g-mode (g$_1$-mode) may become similar to or greater than that of the i-mode, resulting in these modes displaying mixed qualitative features. A similar issue exists for the s$_1$-mode in models with high shear modulus near the crust-core transition relative to the rest of the crust. In such cases, there is not a clearly distinct single i-mode. However, this occurs for a small fraction of our nucleonic model samples, so how we handle qualitatively-mixed modes does not significantly affect our results; we treat the lower-frequency mode as the i-mode.

\subsection{Fitting formula}

The fits for i-mode frequency ($f_{2i}$) in Fig.~\ref{fig:frequency_fit} use NS mass ($M_{\rm NS}$), NS radius ($R_{\rm NS}$) and  
\begin{align}
    N'=\frac{3}{\pi R_{cc}^3}\iiint_0^{R_{cc}}\frac{R_{cc}-r}{R_{cc}}N^2\;d^3r
\end{align}
\noindent where $R_{cc}$ is the crust-core transition radius and $N$ is the Brunt frequency (the frequency of oscillations due to buoyancy, see e.g. \citet{reisenegger_new_1992}). We use this function because we find that it gives better fits for the i-mode frequency than various other functions of the Brunt frequency, although for some other functions the difference is small. 

We explore fitting formulae of the form 
\begin{align}
    f_{2i}={\rm f}_1(M_{\rm NS},R_{\rm NS},N')+{\rm f}_2(M_{\rm NS},R_{\rm NS},N')+\ldots.
\end{align}
\noindent Each term in these formulae is a product of integer powers of $M_{\rm NS}$, $R_{\rm NS}$ and $N'$, and we explore powers in the range -3 to 3 (inclusive). We identify the combinations of two, three and four terms that result in the best fits for the i-mode frequencies of a set of NS models that share the same nuclear matter EOS but that have different inner cores and NS masses, using around 2500 samples from each of our core types and sampling NS mass between 1 and 3 $M_{\odot}$ (or the NS maximum mass if it is lower for the sampled core). This is repeated several times for different choices of the nuclear matter EOS to find the formula that is, on average, the best fit for all of them. The four-term fits are significantly better than those using two or three terms. This leads us to the formula 
\begin{align}
    f_{2i}=a\frac{1}{R_{\rm NS}}+b\frac{M_{\rm NS}}{R_{\rm NS}^2}+c\frac{N'}{R_{\rm NS}}+d\frac{N'M_{\rm NS}^2}{R_{\rm NS}}
\end{align}
\noindent where $a$, $b$, $c$ and $d$ are coefficients to be fit for a choice of nuclear matter EOS. Note that this is not a `universal relation’ for the i-mode frequency, as the coefficients vary significantly for different nuclear matter EOSs. Also note that several formulae using different terms give similar-quality fits and so one should not draw too much meaning from these terms, although $1/R_{\rm NS}$ and $M_{\rm NS}/R_{\rm NS}^2$ are the two most significant terms in most good formulae and have relatively simple physical explanations; the inverse of radius directly affects the period of a wave travelling around the star, and $M_{\rm NS}/R_{\rm NS}^2$ is surface gravity. 

We repeat this fitting process using only NS mass and radius, maintaining that the fits contain 4 terms. The resulting optimal fit is 
\begin{align}
    f_{2i}=a\frac{1}{M_{\rm NS}R_{\rm NS}^3}+bR_{\rm NS}+cR_{\rm NS}^2+d\frac{M_{\rm NS}^3}{R_{\rm NS}^3}
\end{align}
The fit using only NS mass meanwhile is simply a 3rd-order polynomial. 

As shown in Fig.~\ref{fig:frequency_fit}, these fits are accurate for all our parametric core model constructions, indicating that the relationship between i-mode frequency and NS mass and radius is not just a consequence of how the core is constructed. We can therefore reliably use just one of these core model constructions to examine this relationship, so long as it explores a wide range of NS core properties. We choose the FOPT to linear $c_s^2$ model, as it explores the widest ranges for NS radius and i-mode frequency of any of our models for any given NS mass.

\section{Nuclear matter constraints}\label{sec:appendix_NMEOS}

\subsection{$\chi$EFT predictions and their use with the NS meta-model}
The quantifications of theoretical uncertainties in the two works \cite{lim_neutron_2018,drischler_quantifying_2020} we draw $\chi$EFT constraints from are complementary. \citet{lim_neutron_2018} considers a wide range of chiral nucleon-nucleon and three-nucleon interactions at different chiral orders (next-to-next-to-leading and next-to-next-to-next-to-leading order, N2LO and N3LO), with varying cutoffs of momentum in the regulator functions that suppress high-momentum contributions in the chiral interactions. This approach probes higher-order contact interactions and thus the convergence of $\chi$EFT. The calculation in \citet{lim_neutron_2018} uses sophisticated many-body perturbation theory (MBPT) calculations. On the other hand, \citet{drischler_quantifying_2020} uses a Gaussian process-based method with physics-informed hyperpriors to learn the EFT convergence pattern from microscopic EOS calculations order-by-order in the chiral expansion. From the learned convergence pattern, the correlated to-all-orders EFT truncation error can then be directly inferred using the BUQEYE EFT truncation error model \cite{drischler_how_2020,melendez_quantifying_2019}. The EFT truncation characterizes the theoretical error that arises from truncating the EFT expansion in practice at a finite order. \citet{drischler_quantifying_2020} also uses sophisticated MBPT calculations and includes three-nucleon forces up to N3LO \citet{drischler_chiral_2019}, at the same order and with the same momentum cutoff of 500 MeV as the corresponding nucleon-nucleon interactions. Both ab initio calculations with theoretical uncertainties rigorously quantified, taken together, provide a comprehensive uncertainty quantification of the nuclear EOS, both in the limits of PNM and SNM. While the (quantified) EFT truncation error is probably dominant in the density regime of interest in this work, we emphasize that future work may also include uncertainties in the low-energy couplings of the nuclear interactions, which are typically fit to experimental scattering data and few-body observables \cite{epelbaum_high-precision_2020,machleidt_chiral_2011}, and those due to approximations in the solver of the many-body Schr\"odinger equation (here, MBPT only). For this task, developing and applying novel nuclear many-body emulators will be critical \cite{duguet_rooting_2023,armstrong_constraining_2026}. 

As we describe the nuclear matter EOS with expansions around SNM and $n_s$, we take derivatives of each sample drawn from $\chi$EFT to find the expansion coefficients. The truncation of these expansions means that they do not perfectly reproduce the $\chi$EFT samples, but the discrepancies are negligible below $n_s$ and only become significant far above it. The main consequence is instead that the values we obtain for the highest-order coefficients ($Z_0$ and $Z_{\rm sym}$) often cause NS models to become unstable at high densities. As we attach parametric inner cores at a few times $n_s$, which is where these coefficients begin to have a significant impact on the nuclear matter EOS, the details of how we handle this have little effect on our results. We draw $Z_0$ and $Z_{\rm sym}$ from normal distributions that result in stable models: $Z_0=1306\pm214\text{ MeV}$ and $Z_{\rm sym}=-2317\pm379\text{ MeV}$ \cite{margueron_equation_2018}.

\subsection{Experimental nuclear matter constraints}
We primarily use the symmetric and asymmetric matter constraints listed in Table 1 of \citet{tsang_determination_2024}, treating each constraint as an independent normal distribution with the uncertainty given by \citet{tsang_determination_2024} as the 1$\sigma$ value. We also include a constraint based on the calculations of \citet{schwenk_resonant_2005} for a unitary neutron gas, $E_{\rm PNM}(0.0085\text{ fm}^{-3})=2.912\pm0.522\text{ MeV}$, which helps to avoid the extremely soft PNM EOSs preferred by our model construction. We do not use the constraints of \citet{tsang_determination_2024} on the pressure of SNM, as they favour low values above saturation, resulting in many models displaying a softening of the NS EOS at high densities that can produce a peak in buoyancy in the outer core. Such a peak qualitatively changes the behaviour of the i-mode, trapping it in the outer core. While interesting, we leave exploration of the dependence of the i-mode's qualitative behaviour on nuclear matter to future work and limit this work to considering stiffer cores that display more typical i-mode behaviour. Such cores are compatible with pressure constraints from other experiments. Based on \citet{oliinychenko_sensitivity_2023}, which analysed recent proton flow data from the STAR experiment, we use $P_{\rm SNM}(2n_s)=19.36\pm7.30\text{ MeV/fm}^3$, $P_{\rm SNM}(3n_s)=109.08\pm18.10\text{ MeV/fm}^3$ and $P_{\rm SNM}(4n_s)=104.91\pm25.07\text{ MeV/fm}^3$. 

As our nucleonic matter modelling is similar to that of \citet{tsang_determination_2024}, we adopt their priors, including the fixed values of $E_0=-16\text{ MeV}$ and $n_s=0.16\text{ fm}^{-3}$. We also use their bounds on the pressure of high-density symmetric nuclear matter $P_{\rm SNM}(4n_s)<300\text{ MeV/fm}^3$, which, for a nuclear model truncated at fifth-order in density, is effectively a $Q_0$-dependent bound on $Z_0$. All of the constraints we use are shown in the supplementary material, alongside the distribution for the nuclear matter EOS that results from applying them to the priors. 

To identify the additional nuclear matter constraints that would most improve the radius posteriors inferred from a measurement of the i-mode frequency, we take a large set of nuclear matter samples that follow these constraints and accept or reject them based on new constraints on the EOS of SNM, the EOS of PNM or the symmetry energy at densities ranging from $0.25n_s$ to $3n_s$. These new constraints are constructed such that the accepted samples follow a normal distribution at the chosen density, with the same mean as the current samples but with the standard deviation reduced by a factor of $1/4$ or $1/16$. We find that for a single $1/4$ constraint, reducing uncertainty in the symmetry energy around $0.19\text{ fm}^{-3}$ is the most effective for strengthening radius constraints, tightening the standard deviation of posterior radius samples for the injection $f_{2i}=325\pm3.25\text{ Hz}$ from $\sigma_R\approx0.72\text{ km}$ to $\sigma_R\approx0.63\text{ km}$. But we also find that constraining the symmetry energy at both $0.12$ and $0.32\text{ fm}^{-3}$ is significantly more effective, with $1/4$ reductions in both uncertainties resulting in a radius posterior with $\sigma_R\approx0.52\text{ km}$. Extending this to three or four new constraints meanwhile results in smaller improvements to the radius ($\sigma_R\approx0.49\text{ km}$ and $\sigma_R\approx0.46\text{ km}$, respectively). We therefore choose to focus on the effect of simultaneously adding constraints on the symmetry energy at $0.12$ and $0.32\text{ fm}^{-3}$. We caution that the construction of nuclear matter in the meta-model we use creates correlations in the EOS across densities, including correlations between the EOS above and below nuclear saturation density, so different methods for constructing the EOS may favour different constraints to these. 

For corner plots showing the nuclear matter parameter distributions resulting from both sets of $\chi$EFT constraints, experimental constraints, and all of their improvements, see supplementary material.

\bibliography{paper6}

\end{document}


\title{Supplementary Material}

\maketitle

\renewcommand{\thefigure}{S\arabic{figure}}
\renewcommand{\thetable}{S\arabic{table}}

\section{NS model construction}
We construct spherically symmetric equilibrium NS models following the NS metamodelling approach of \citet{margueron_equation_2018}, \citet{grams_properties_2022} and \citet{margueron_equation_2018-1}. We neglect effects of rotation, magnetic fields and finite temperatures under the assumptions that NSs involved in binary mergers are old and that tidal heating prior to resonant i-mode excitation is not significant \cite{lai_resonant_1994}.

\subsection{Equation of state of bulk nucleonic matter}

We begin with models for the EOS of dense nuclear matter of the form: 
\begin{align}
    E(n,\delta)=E(n,0)+E_{\rm sym}(n)\delta^2+\mathcal{O}(\delta^4),
\end{align}
\noindent where $n$ is baryon density and $\delta=1-2y_p$ is isospin asymmetry, with $y_p$ being the proton fraction. Symmetric nuclear matter (SNM) is given by $\delta=0$ and pure neutron matter (PNM) by $\delta=1$, which are separated by the nuclear symmetry energy ($E_{\rm sym}$). These are further expanded in density as 
\begin{align}
    E(n,0)=E_0+\frac{1}{2}K_0\chi^2+\frac{1}{3!}Q_0\chi^3+\frac{1}{4!}Z_0\chi^4+\mathcal{O}(\chi^5)
\end{align}
\begin{align}\nonumber
    E_{\rm sym}(n)=J+&L\chi+\frac{1}{2}K_{\rm sym}\chi^2\\&+\frac{1}{3!}Q_{\rm sym}\chi^3+\frac{1}{4!}Z_{\rm sym}\chi^4+\mathcal{O}(\chi^5),
\end{align}
\noindent with $\chi=(n-n_s)/(3n_s)$ and $n_s$ being the nuclear saturation density (the density at which $\partial E(n,0)/\partial n=0$). With the truncation used here, this model for bulk nucleonic matter has 10 free parameters: $E_0$, $n_s$, $K_0$, $Q_0$ and $Z_0$ (isoscalar parameters), and $J$, $L$, $K_{\rm sym}$, $Q_{\rm sym}$ and $Z_{\rm sym}$ (isovector parameters). We also add an exponential term that ensures that the limit $E(0,\delta)=0$ is satisfied while keeping the EOS smooth (see section III-D of \citet{margueron_equation_2018}; as in that work, we use $b=10\ln2$). This term decreases with increasing density such that it is negligibly small near the NS crust-core transition.

\subsection{Nucleonic NS component – crust and outer core}

The nuclear matter EOS determines bulk contributions to the properties of nucleonic NS matter and thus plays a key role in determining its equilibrium structure, which is qualitatively different at high and low densities: the NS core consists of uniform matter, while the crust is a non-uniform system of nuclear clusters immersed (at densities beyond neutron drip) in a gas of free neutrons. We model the crust with a Compressible Liquid Drop Model (CLDM), wherein a spherical unit cell contains a uniform spherical cluster surrounded by uniform free neutrons. All clusters are assumed to be identical and arranged in a simple-cubic crystal lattice. The surface and curvature contributions arising from the interface between the clusters and neutron gas are given by a function of the proton fraction with five free parameters which, for simplicity, we fix to the `standard’ values given in \citet{grams_properties_2022}: $\sigma_{\rm surf}=1.1 \text{ MeV/fm}^2$, $b_{\rm surf}=29.9$, $p_{\rm surf}=3.0$, $\sigma_{\rm curv}=0.1 \text{ MeV/fm}$ and $\beta_{\rm curv}=0.7$. The crust-core transition typically occurs around half the nuclear saturation density. 

The equilibrium volume fraction of nuclear clusters ($v$) increases throughout the NS crust, reaching a significant fraction of unity near the crust-core transition. Where this occurs the assumption that clusters are spherical breaks down, as complicated nuclear geometries known as `nuclear pastas' are predicted to appear \cite{ravenhall_structure_1983,lorenz_neutron_1993}. Self-consistently modelling the appearance and properties of pasta phases is beyond the scope of this work, so we instead follow \citet{pethick_matter_1995} and estimate that a transition from spherical clusters to nuclear pastas occurs where $v=1/8$. As nuclear pastas have significantly different elastic properties to spherical clusters \cite{pethick_liquid_1998}, we replace the shear modulus beyond this transition with the smooth cubic approximation used by \citet{sotani_constraints_2011}. We also linearly extrapolate the energies per baryon of uniform and spherically clustered matter at $v=1/8$ to the baryon density at which they cross, and use that as the crust-core transition density instead of the value obtained from using spherical clusters at $v>1/8$. At densities below the crust-core transition we continue to use the pressure and buoyancy obtained for spherical clusters, as these are much less affected by the appearance of pastas than elasticity (see, e.g., figures 36-38 of \citet{chamel_physics_2008}). 

The following are crust model assumptions/approximations and their impacts. 
\begin{itemize}
    \item Simple cubic lattice: The corrections for other lattice types, such as bcc and fcc, are at the level of 1\%.  
    \item One component crust: there are suggestions that the crust contains a non-negligible fraction of impurities from neutron star surface cooling measurements in qLMXBs. A multi-component crust will not significantly affect the equation of state, but could affect the shear modulus.  
    \item Fixed surface and curvature parameters: varying these parameters over ranges given by fitting the CLDM to nuclear masses and microscopic mean-field calculations \cite{carreau_crystallization_2020,grams_neutron_2025} causes i-mode frequencies to vary by less than one Hertz, and NS radii by at most tens of meters. Such effects are small in the scope of this work. 
    \item  Nuclear pasta approximation: fully self-consistently calculating pasta phases may result in additional i-modes appearing at the boundary at which pasta first appears, but these i-modes are likely to have similar dependences on NS radius as crust-core i-modes. 
\end{itemize}

\subsection{Non-nucleonic inner cores}

\begin{figure}
    \centering
    \includegraphics[width=\columnwidth]{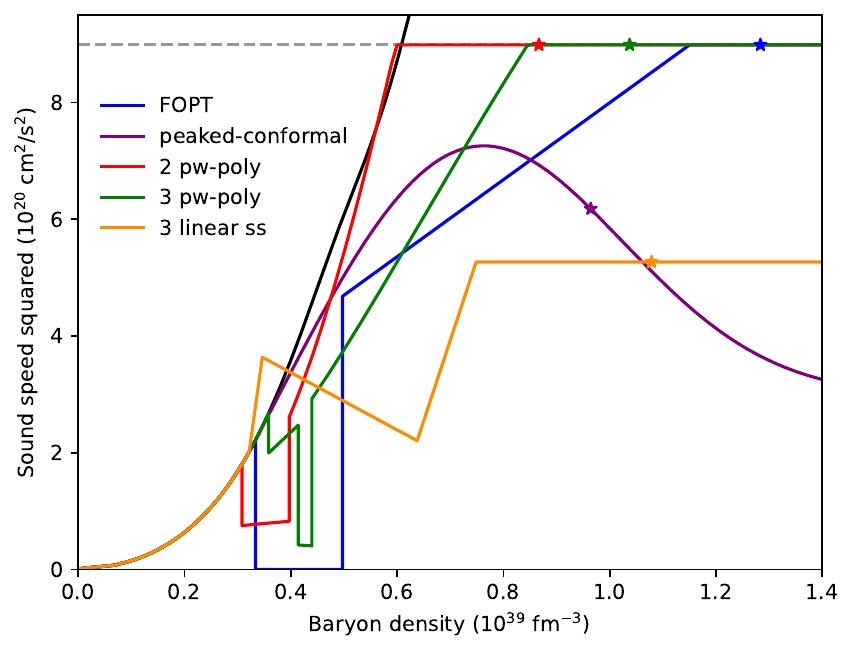}
    \caption{Examples of the parametric core models used to test our fits for the i-mode frequency in Fig.~1. The stars indicate the central density of the maximum mass NSs for each of these examples. \textbf{Black:} the original nucleonic model. \textbf{Blue:} FOPT to linear $c_s^2$ model. \textbf{Purple:} peaked-conformal model. \textbf{Red:} two piecewise-segment polytropic model. \textbf{Green:} three piecewise-segment polytropic model. \textbf{Orange:} three-segment linear $c_s^2$ model.}
    \label{fig:cores}
\end{figure}

Exotic degrees of freedom may become significant in the NS inner core, causing its properties to deviate from what is predicted for nucleonic matter. To allow for this additional freedom, we replace nucleonic NS EOSs beyond some baryon density $n_{t1}$ (which is a free model parameter) with realisations of parametric inner-core models. Equations for these models and the prior ranges we use for their free parameters are given in Table~\ref{tab:table1}, and an example realisation of each model is shown in Fig.~\ref{fig:cores}. The first model we consider features a first-order phase transition (FOPT), followed by sound speed squared ($c_s^2$) increasing linearly with baryon density. The FOPT allows for this model to explore a wide range of inner-core properties with minimal influence from the nucleonic part of the NS above it. We also use piecewise-polytropic models (similar to \citet{vuille_maximum_1999,read_constraints_2009}) with either two or three polytropic segments, and models where $c_s^2$ varies linearly with baryon density in each of three density segments in the core (similar to \citet{somasundaram_investigating_2023}). These models give continuous EOSs, but have more free parameters than the FOPT model, allowing a significant but somewhat different space of inner cores to be explored. Finally, we consider NS cores constructed following \citet{tews_constraining_2018} that feature a peak in sound speed before trending towards the conformal limit of perturbative quantum chromodynamics (i.e., $c_s^2$ approaches $c^2 /3$ at extreme densities, where $c$ is the speed of light), which we refer to as the `peaked-conformal’ model. We require that the peak in sound speed be above rise above $c^2 /3$ (rejecting samples that do not, or that peak above $c^2$) and then dip back down below it, such that the conformal limit is approached from below. This model has $c_s^2$ and its baryon density derivative be continuous, and thus the nucleonic NS model to which it is attached has a greater influence on the distribution of inner-core properties this model produces than for the other core models. For all models, we reject samples that result in maximum NS masses below $2M_{\odot}$, following observational constraints \cite{fonseca_refined_2021}.

\begin{table*}
\caption{\label{tab:table1}The different types of core models we attach to nucleonic NS EOSs in this work to explore the connection between i-mode frequency and NS radius, and the ranges within which we sample their parameters. The prior parameter ranges are further modified by an observational constraint \cite{fonseca_refined_2021} on the NS maximum mass, $M_{\rm max}>2.0M_{\odot}$. After constructing the NS EOS we sample mass in the range $1<M_{\rm NS}/M_{\odot}<3$ and solve the TOV equations to obtain a single NS model, rejecting samples with $M_{\rm NS}>M_{\rm max}$. We also reject NS models where the transition density $n_{t1}$ is not reached, which favours low transition densities, with around 30\% of accepted samples for any core model having $n_{t1}>0.32\text{ fm}^{-3}$ and only 5\% having $n_{t1}>0.48\text{ fm}^{-3}$. Note that the square of the sound speed $c_s^2(n)$ and the NS EOS $P(\rho)$ are related by $c_s^2=dP/d\rho=(nd\mu)/(\mu dn)$, where $P$ is pressure, $\rho$ is energy density and $\mu$ is the chemical potential. The density at which $c_s^2$ reaches $c^2$ is referred to as $n_{\rm acausal}$. For the polytropic models, $K_{1-3}$ are set such that the NS EOS is continuous (including continuity with the nucleonic NS EOS model at $n_{t1}$), as is $c_s^2(n_{t1})$ for the linear model. $c_1$ and $c_2$ for the peaked-conformal model are set such that both the EOS and its slope in baryon density are continuous at $n_{t1}$, with $n_{t1}$ being a free parameter.}
\begin{ruledtabular}
\begin{tabular}{ccc}
 & NS inner core EOS model & Prior parameter ranges \\ \hline
FOPT to linear $c_s^2$ & $c_s^2(n)=\begin{cases}0,&n_{t1}\leq n< n_{t1}+\Delta n\\c_{\rm pt}^2+(n-n_{t1}-\Delta n)dc_{pt}^2,&n_{t1}+\Delta n\leq n< n_{\rm acausal}\\c^2,&n_{\rm acausal}\leq n\end{cases}$ & $\begin{array}{l} 0.24<n_{t1}<0.80\text{ fm}^{-3} \\ 0<\Delta n<0.48\text{ fm}^{-3} \\ 0<c_{\rm pt}^2<c^2 \\ -2<\log \left(\frac{dc_{\rm pt}^2}{c^2\; \text{fm}^3}\right)<4 \end{array}$ \\ \hline
2 piecewise-polytrope & $P(\rho)=\begin{cases}K_1\rho^{\gamma_1},&n_{t1}\leq n\leq n_{t2}\\K_2\rho^{\gamma_2},&n_{t2}\leq n\leq n_{t3}\\K_3\rho^{\gamma_3},&n_{t3}\leq n\leq n_{\rm acausal}\\P(n_{\rm acausal})+c^2\rho,&n_{\rm acausal}\leq n \end{cases}$ & $\begin{array}{l} 0.24<n_{t1-t2}<0.80\text{ fm}^{-3} \\ 0<\gamma_{1-2}<5 \end{array}$ \\ \hline
3 piecewise-polytrope & $P(\rho)=\begin{cases}K_1\rho^{\gamma_1},&n_{t1}\leq n\leq n_{t2}\\K_2\rho^{\gamma_2},&n_{t2}\leq n\leq n_{\rm acausal}\\P(n_{\rm acausal})+c^2\rho,&n_{\rm acausal}\leq n \end{cases}$ & $\begin{array}{l} 0.24<n_{t1-t3}<0.80\text{ fm}^{-3} \\ 0<\gamma_{1-3}<5 \end{array}$ \\ \hline
3 linear $c_s^2$ & $c_s^2(n)=\begin{cases}c_s^2(n_{t1})+\frac{n-n_{t1}}{n_{t2}-n_{t1}}(c_s^2(n_{t2})-c_s^2(n_{t1})),&n_{t1}\leq n\leq n_{t2}\\c_s^2(n_{t2})+\frac{n-n_{t2}}{n_{t3}-n_{t2}}(c_s^2(n_{t3})-c_s^2(n_{t2})),&n_{t2}\leq n\leq n_{t3}\\c_s^2(n_{t3})+\frac{n-n_{t3}}{n_{t4}-n_{t3}}(c_s^2(n_{t4})-c_s^2(n_{t3})),&n_{t3}\leq n\leq n_{t4}\\c_s^2(n_{t4}),&n_{t4}\leq n\end{cases}$ & $\begin{array}{l} 0.24<n_{t1-t4}<0.80\text{ fm}^{-3} \\ 0<c_s^2(n_{t2-t4})<c^2 \end{array}$ \\ \hline
peaked-conformal & $c_s^2=\frac{1}{3}-c_1\exp\left(-\frac{(n-c_2)^2}{n_{\rm BL}^2}\right)+h_p\exp\left(-\frac{(n-n_p)^2}{w_p^2}\right)\left(1+\text{erf}\left(s_p\frac{n-n_p}{w_p}\right)\right)$ & $\begin{array}{l} 0.24<n_{t1}<0.48\text{ fm}^{-3} \\ 0.01<n_{\rm BL}<3.20\text{ fm}^{-3} \\ 0<h_p<1 \\ n_{t1}<n_p<3.20\text{ fm}^{-3} \\ 0.08<w_p<3.20\text{ fm}^{-3} \\ -50<s_p<50 \end{array}$ \\ 
\end{tabular}
\end{ruledtabular}
\end{table*}

In addition to the NS EOS, we must also choose descriptions for compositional properties of the inner core that affect i-mode oscillations: buoyancy and elasticity. For elasticity, some possible exotic phases are solid and thus allow shear forces \cite{pandharipande_model_1975,owen_maximum_2005}. However, the presence of an interface at which shear forces become non-zero in the core would likely not significantly affect crust-core interface modes, but rather would contribute to the appearance of a separate family of core interface modes. We therefore expect that allowing elastic inner cores would not significantly alter the results of this work, and so keep the shear modulus at zero in the inner core. Buoyancy driven by compositional gradients meanwhile is effectively suppressed by equilibrating processes (e.g., beta decay), which damp oscillations \cite{reisenegger_new_1992}. These processes are slow in cold nucleonic NS matter (relative to the oscillations of interest to this work), but in exotic phases of matter they may be significantly faster \cite{baym_neutron_1995}. We therefore assume that exotic matter equilibrates on timescales much shorter than mode oscillations. While some of the core models we use feature smooth EOSs at $n_{t1}$, for simplicity we discontinuously transition to zero buoyancy at densities above $n_{t1}$ for all models. This is a significant assumption, so more detailed examinations of how compositional buoyancy is affected by non-nucleonic degrees of freedom at high densities are a key area for future study.

\subsection{Notes of the impact of buoyancy on this work}

The method presented in this work is most reliable if buoyancy’s impact on the i-mode's frequency is low, leaving bulk radius and mass as the i-mode's only core dependences. For the fixed nucleonic EOS we used for Fig.~1, this is the case, and, as stated above, our core model construction assumes that there is effectively no buoyancy in the inner core due to exotic matter having fast equilibrating reactions \cite{baym_neutron_1995}. Note that we focus entirely on compositional buoyancy (which arises from matter stratification), assuming zero temperature and thus no thermal buoyancy. Significant temperatures would be required for thermal buoyancy to be important \cite{gittins_neutron-star_2025}, which is unlikely prior to binary coalescence \cite{lai_resonant_1994}. 

\begin{figure*}
    \centering
    \includegraphics[width=\textwidth]{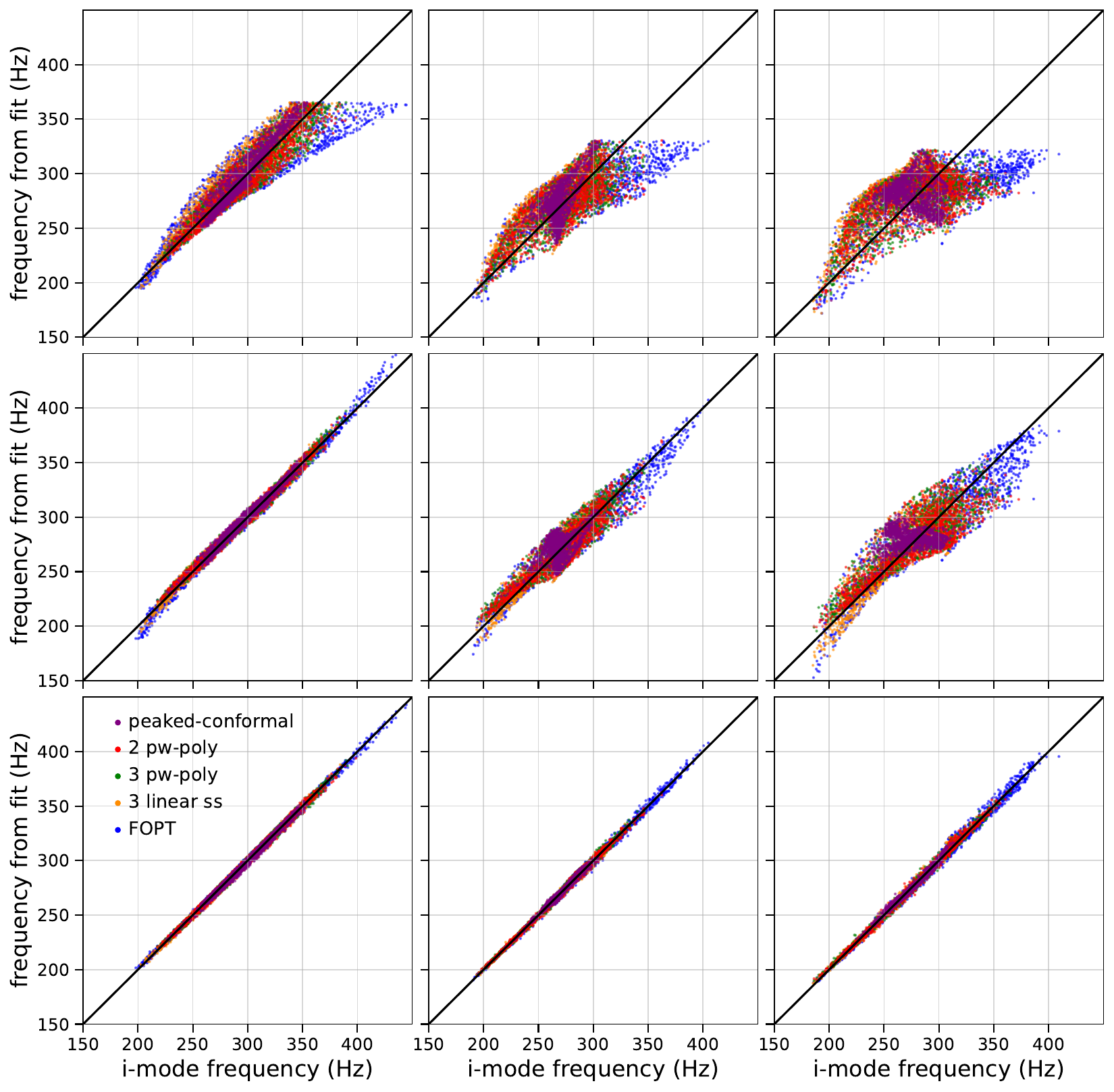}
    \caption{Fits to the i-mode frequency, using only mass (top), mass and radius (middle, or mass, radius and a weighted average of the compositional buoyancy (bottom), similar to Fig.~1. The first column is the same as in Fig.~1, and the other two are for different models for the fixed low-density nucleonic physics. The more stratified the NS model is, the more important buoyancy is to the fits.}
    \label{fig:other_fits}
\end{figure*}

For more stratified nucleonic models than the one used for Fig.~1, compositional buoyancy is larger in the NS crust and outer core. In such cases, the integrated function of buoyancy we use in our fitting formulae (see Appendix~A) is more important for accurately fitting the i-mode frequency. This can be seen in Fig.~\ref{fig:other_fits}, where, for a strongly stratified nucleonic model (right column), not considering buoyancy effectively adds 50 Hz uncertainty to the i-mode frequency. This is however accounted for in our results, as for Fig.~2 we allow reasonable ranges of nucleonic models without any additional constraints on buoyancy outside the inner core. We are able to recover NS radius with reasonable precision, indicating that, while buoyancy strongly affect the i-mode frequency of some NS models, for most models the radius is more significant. 

For the inner core, however, we have only considered models with zero effective buoyancy. If a model were used in which buoyancy did not discontinuously become zero (as may be more appropriate for, e.g., the continuous cores produced by the `peaked-conformal’ model), the i-mode may penetrate further into the core, resulting in more localised dependences on core properties. This would result in radii inferred from the i-mode frequency having some systematic dependence on the modelling used for the NS inner core. Buoyancy in the NS core may however yet be constrained, as observations of GW signals with next-generation GW facilities may allow us to constrain properties of multiple gravitational-modes (g-modes). The g-modes are most strongly dependent on buoyancy in the NS core, with different modes in that family being sensitive to different regions of the core. Detecting several g-modes could therefore provide detailed insight into the buoyancy profile of the NS core, allowing us to determine the significance of inner-core buoyancy. 

We find that the methods described in this work are reliable under reasonable assumptions, but may require additional information from g-modes if buoyancy is significant in the NS inner core. The assumption that the core has zero effective buoyancy is significant, and any real observations should account for the possibility of other behaviours. However, this assumption is not unreasonable, particularly for cores featuring first-order phase transitions.

\section{Results Table and Figures expanding on the main text}

Table~\ref{tab:table2} is a summary of our priors (informed by nuclear matter EOS constraints and the NS mass measurement) and posteriors (also informed by i-mode frequency measurements) for NS radius. NS radius may be recovered to within 0.908 km if uncertainties in nucleonic physics are negligible and i-mode frequency is well-constrained, but with current constraints the posteriors are significantly wider. However, even for the most pessimistic scenario examined in this work, we still see some improvement over the prior (from 4.247 to 3.707 km for $Chi$EFT constraints from \citet{lim_neutron_2018} and a 5\% frequency uncertainty). Note also that a more reasonable `pessimistic' scenario for analysing real data would combine constraints from experiment and theory, rather than using only one or the other as we have done in this work, which would be give stronger constraints.

\begin{table*}
\caption{\label{tab:table2}Summary of the priors and posteriors for NS radius recovery with different injected i-mode frequency measurements and constraints on dense nuclear matter, some of which are shown in Fig.~2. The values given are minimum 90\% confidence intervals for NS radius, in kilometres. The columns are for different constraints on the EOS of dense nuclear matter, and rows are for uncertainties in the i-mode frequency measurement (5\%, 3\% or 1\% of the mean of the injected normal distribution, or the prior for no measurement), and all of these recoveries have NS mass constrained to follow the low-spin posteriors for the lower mass object of GW70817. Note that the radius is injected at 11 km and the mean of the frequency injection is a value approximately at the centre of the prior distribution at that radius, which is different for the different nucleonic constraints (365 Hz for fixed nucleonic physics, 350 Hz for $\sigma_{\rm L\&H}$ and $\sigma_{\rm D}$ and their improvements, and 325 Hz for $\sigma_{\rm exp}$. and its improvements).}
\begin{ruledtabular}
\begin{tabular}{ccccccccccc}
 & fixed nucleonic physics & $\sigma_{\rm L\&H}$ & $\sigma_{\rm L\&H}/4$ & $\sigma_{\rm L\&H}/16$ & $\sigma_{\rm exp}$ & $\sigma_{\rm exp/4}$ & $\sigma_{\rm exp}/16$ & $\sigma_{\rm D}$ & $\sigma_{\rm D}/4$ & $\sigma_{\rm D}/16$ \\ \hline
\vspace{-8pt} \\
Prior & 3.496 & 4.247 & 3.660 & 3.484 & 4.071 & 3.403 & 3.338 & 3.526 & 3.356 & 3.323 \\ 
5\% & 2.441 & 3.707 & 2.961 & 2.607 & 2.626 & 2.280 & 2.219 & 2.939 & 2.620 & 2.457 \\ 
3\% & 1.644 & 3.573 & 2.621 & 2.033 & 2.375 & 1.982 & 1.940 & 2.766 & 2.236 & 1.861 \\ 
1\% & 0.908 & 3.499 & 2.362 & 1.488 & 2.178 & 1.709 & 1.615 & 2.655 & 1.907 & 1.248 \\ 
\end{tabular}
\end{ruledtabular}
\end{table*}

\begin{figure*}
    \centering
    \includegraphics[width=0.67\textwidth]{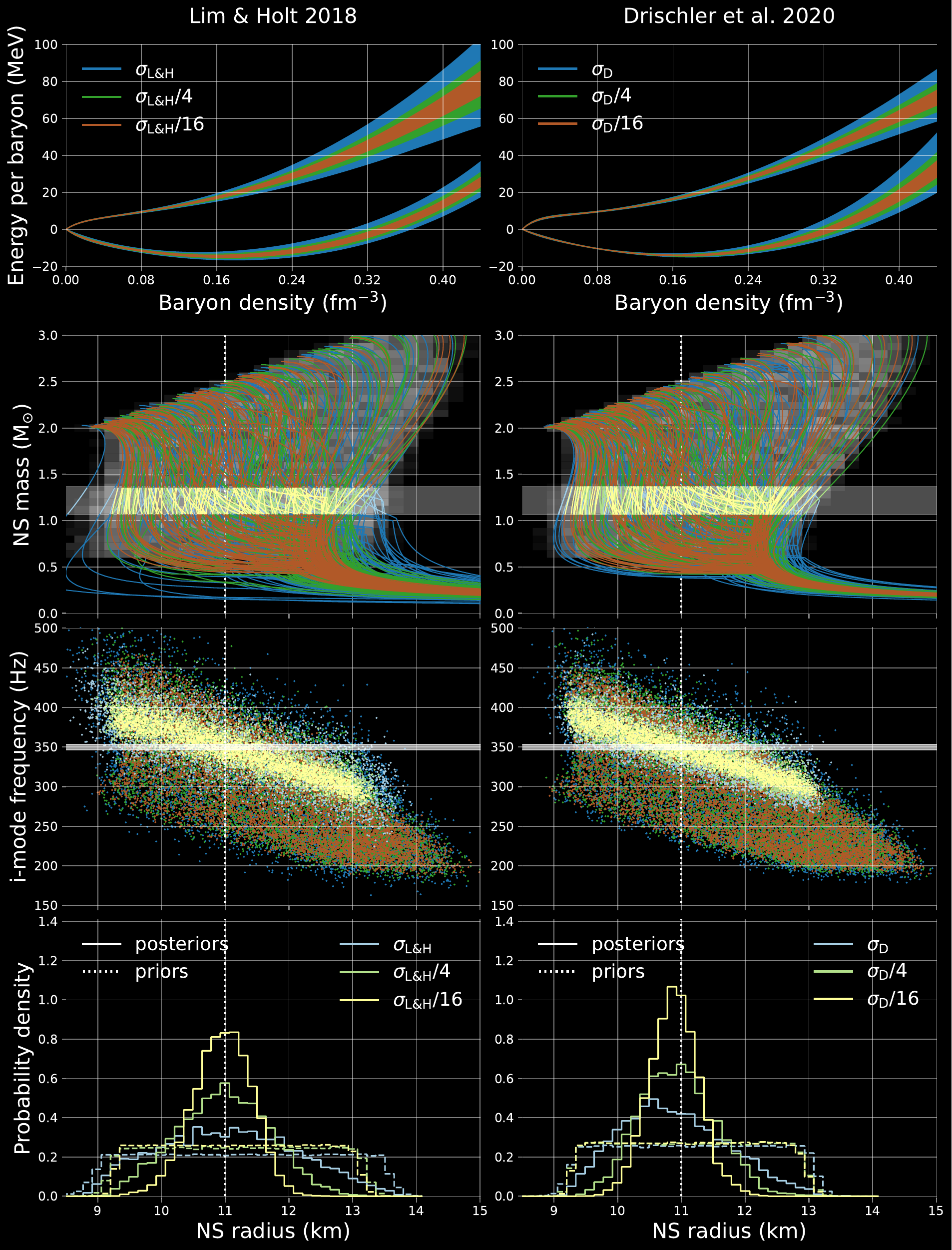}
    \caption{Similar to Fig.~2 in the main text, showing the process of inferring NS radius from i-mode frequency. The top row shows 1$\sigma$ ranges for the EOSs of symmetric nuclear matter and pure neutron matter, and the second row shows mass-radius curves generated by attaching parametric cores to NS models generated from those nuclear matter EOSs. The green bar indicates the span of the posterior for the mass of the lighter object in GW170817, which we take as the mass constraint for this inference. The third row shows radii and frequencies for sampled NSs with masses over a wide range (darker colours) and over the GW170817 posterior range (lighter colours), with a green bar showing the 1$\sigma$ range of the synthetic frequency measurement we inject to obtain the NS radius posteriors shown in the final row. This Figure compares results using nuclear matter constraints taken from \citet{lim_neutron_2018} (i.e. the second column of Fig.~2) and results using constraints from \citet{drischler_quantifying_2020}. We see that the results are qualitatively similar, meaning that $\chi$EFT systematics currently only have a small impact on the method described in this work.}
    \label{fig:inferences_XEFT}
\end{figure*}

Fig.~\ref{fig:inferences_XEFT} is similar to Fig.~2 of the main text, but for the inference of NS radius using nuclear matter constraints informed by \citet{lim_neutron_2018} and \citet{drischler_quantifying_2020} so that they can be easily compared. We see that there are quantitative differences in the dense matter EOS constraints, and that they propagate to the radius posteriors, but the posteriors are qualitatively similar. The constraints using \citet{drischler_quantifying_2020} are somewhat stronger than those using \citet{lim_neutron_2018}, so in the main text we use the latter to be conservative, but the differences have negligible impact on the main findings of this work.

\begin{figure}
    \centering
    \includegraphics[width=\columnwidth]{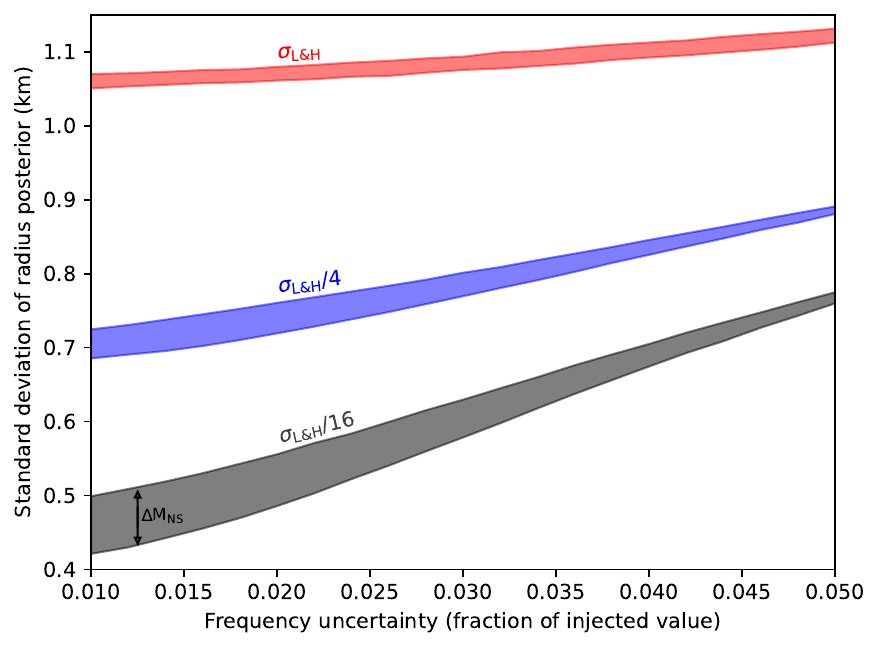}
    \caption{The dependence of the standard deviation of the posterior distribution for NS radius ($\sigma_R$) on the precision of the i-mode frequency measurement, NS mass measurement and uncertainties in nucleonic physics. The different colours are for the current $\chi$EFT predictions of \citet{lim_neutron_2018} (red) and the improvements introduced in the main text (blue for uncertainties reduced by a factor of 1/4, and grey for 1/16). The x-axis is the fractional uncertainty in the i-mode frequency measurement (at 1$\sigma$). The width of each coloured band corresponding to NS mass uncertainty in the range $0.01<\sigma_q<0.09$. For the uncertainty ranges shown, nucleonic uncertainties are dominant, but the precision of the frequency and mass measurements also become important as those uncertainties are reduced.}
    \label{fig:uncertainties}
\end{figure}

Fig.~\ref{fig:uncertainties} provides an overview of the relative impacts of uncertainties in NS mass, i-mode frequency and nucleonic physics on our radius recovery. For nucleonic uncertainties we use the $\chi$EFT predictions of \citet{lim_neutron_2018} and the `improvements' given in the main text, and for i-mode frequency measurement we inject a normal distribution with a 1\%, 3\% or 5\% standard deviation. We inject NS mass uncertainty as a normal distribution in the mass ratio of the binary (truncated at 0 and 1), fixing the chirp mass to the mean value for the low-spin posteriors of GW170817 \cite{abbott_GW170817_2017} so that mass ratio directly corresponds to NS mass (using the less-massive NS, as in the main text). We centre this distribution at $q=0.85$ and vary the standard deviation ($\sigma_q$) between $0.01$ and $0.09$ (for reference, in this setup, the posteriors of GW170817 used in the main text have $\sigma_q\approx0.051$). We see that, over these ranges, nucleonic uncertainties are dominant, hence our focus on them in the main text. As the nucleonic uncertainties are reduced, however, the uncertainties in the measurements of NS mass and i-mode frequency do become more significant, with i-mode frequency uncertainty being somewhat more impactful than NS mass uncertainty. This indicates that while advances in terrestrial nuclear physics are most effective at present, it is important that GW facilities and analysis techniques also keep improving.

\begin{figure*}
    \centering
    \includegraphics[width=\textwidth]{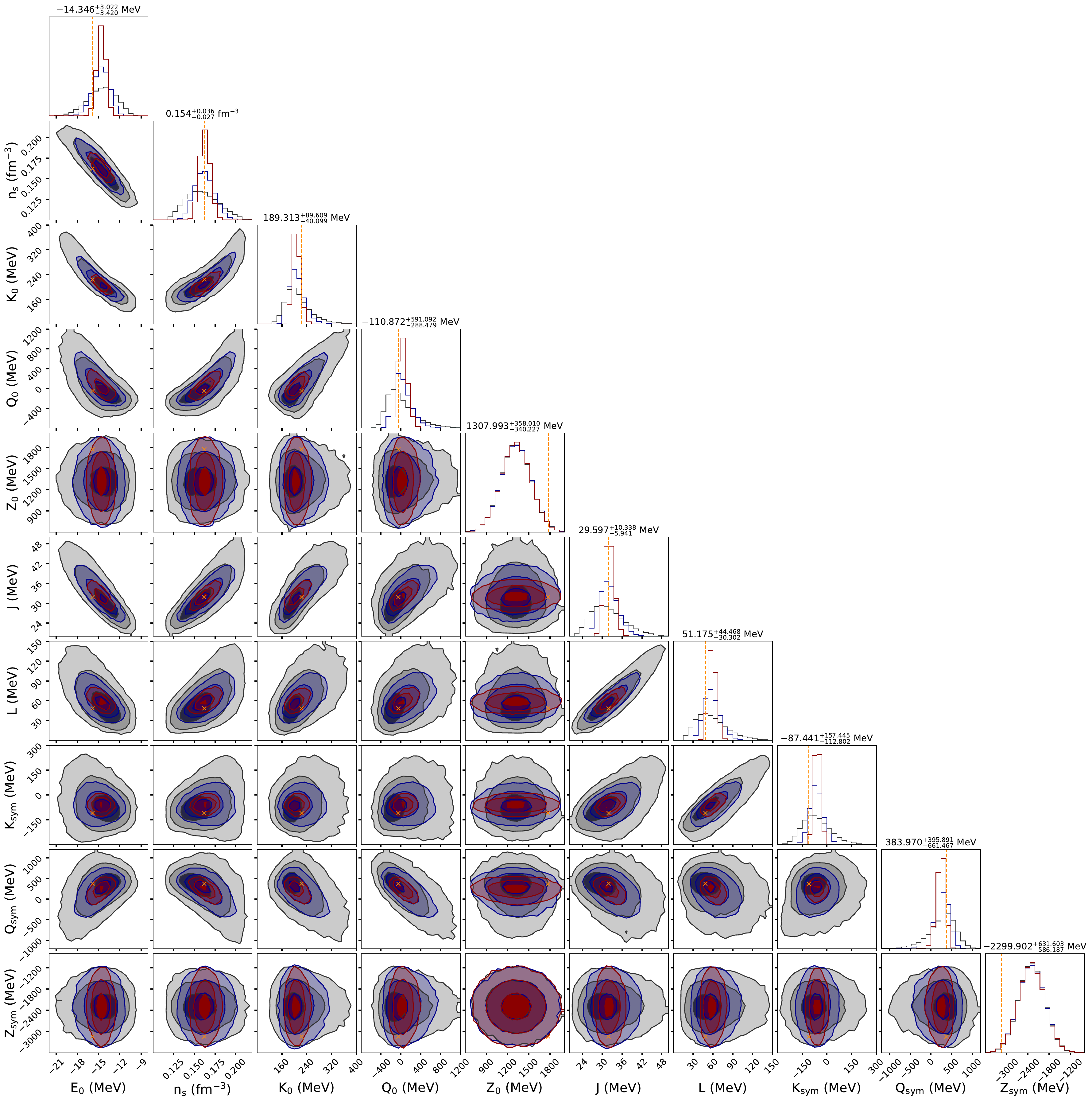}
    \caption{Corner plot showing the NS metamodel parameter distribution for the $\chi$EFT predictions from \citet{lim_neutron_2018}, and the reduced uncertainties we use. Grey is for the constraints from \citet{lim_neutron_2018}, blue is for uncertainties reduced by a factor of 4, and red for a factor of 16. The $Z_0$ and $Z_{\rm sym}$ distributions are modified for compatibility with the NS metamodel we use, and are not changed when the uncertainties are reduced. The orange markers/lines indicate the nucleonic model we use when fixing low-density physics (Fig.~1 and the left column of Fig.~2 in the main text), which is chosen to be a sample consistent with this distribution. Numbers above each column indicate the marginalised 1-D unimproved (grey) constraints on each parameter, written as the location of the peak plus/minus the ranges to the bounds of the minimum 90\% confidence interval.}
    \label{fig:corner_JXEFT}
\end{figure*}

\begin{figure*}
    \centering
    \includegraphics[width=\textwidth]{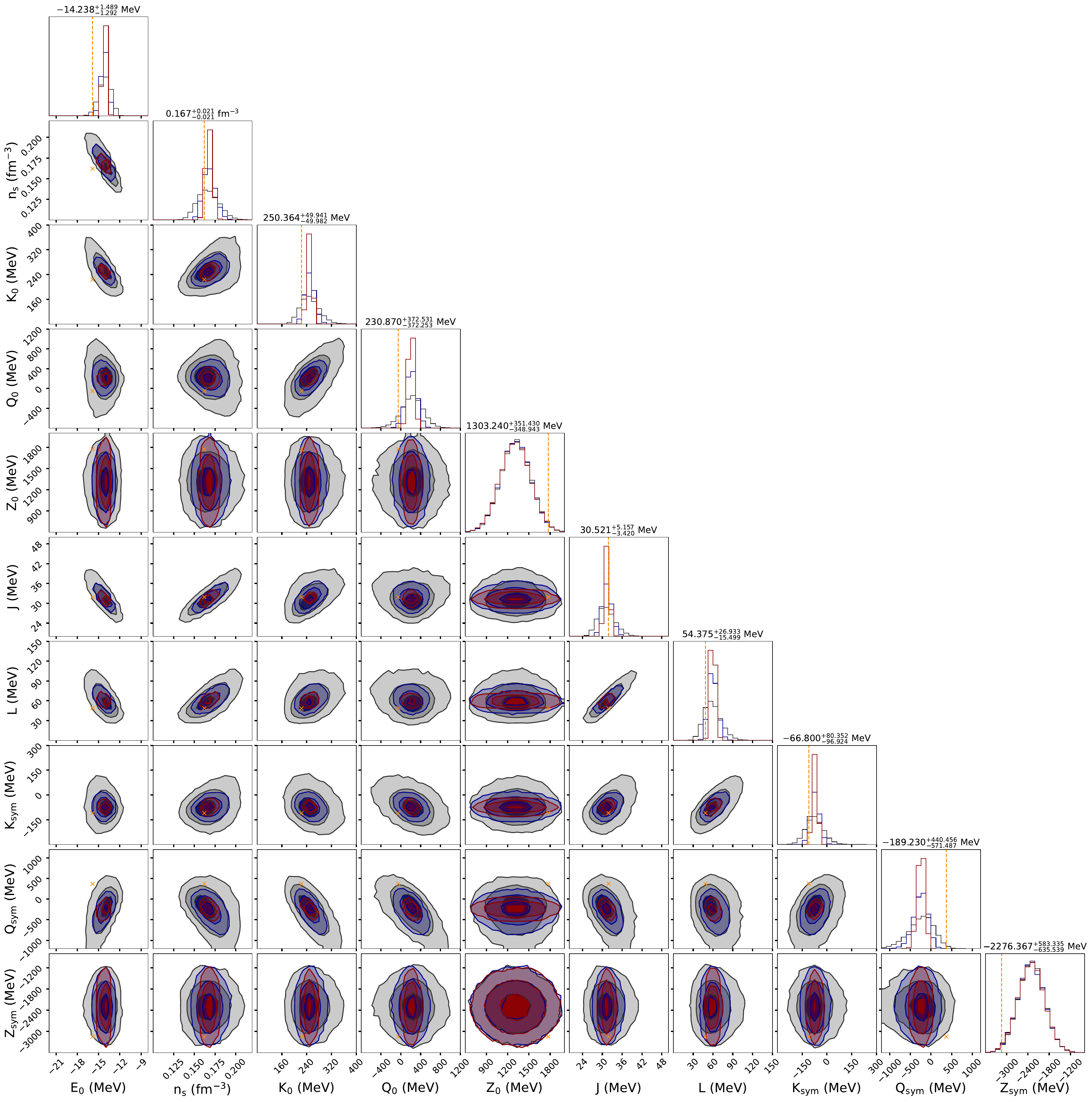}
    \caption{Similar to Fig.~\ref{fig:corner_JXEFT}, but for the $\chi$EFT uncertainties from \citet{drischler_quantifying_2020}. The axis scales are the same as those in Fig.~\ref{fig:corner_JXEFT}.}
    \label{fig:corner_CXEFT}
\end{figure*}

\begin{figure*}
    \centering
    \includegraphics[width=\textwidth]{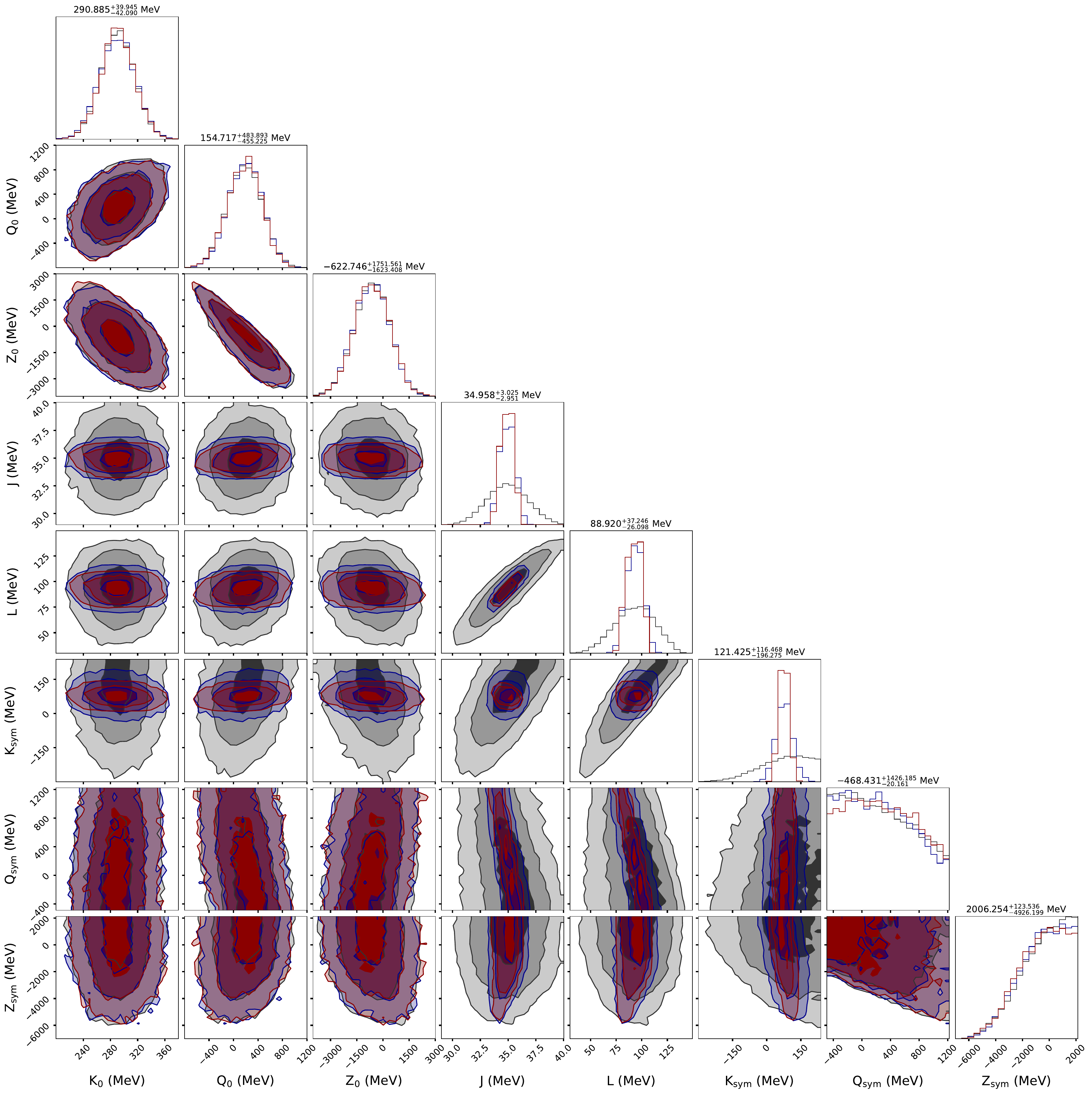}
    \caption{Similar to Fig.~\ref{fig:corner_JXEFT}, but for uncertainties from nuclear experiment. $E_0$ and $n_s$ are not shown, as they are fixed to $-16 \text{ MeV}$ and $0.16\text{ fm}^{-3}$, respectively. Note that the axis scales are different in this Figure to those in Figs.~\ref{fig:corner_JXEFT} and~\ref{fig:corner_CXEFT}, and that the $K_{\rm sym}$, $Q_{\rm sym}$, and $Z_{\rm sym}$ distributions reach the edges of the prior bounds that we use.}
    \label{fig:corner_exp}
\end{figure*}

\begin{figure*}
    \centering
    \includegraphics[width=\textwidth]{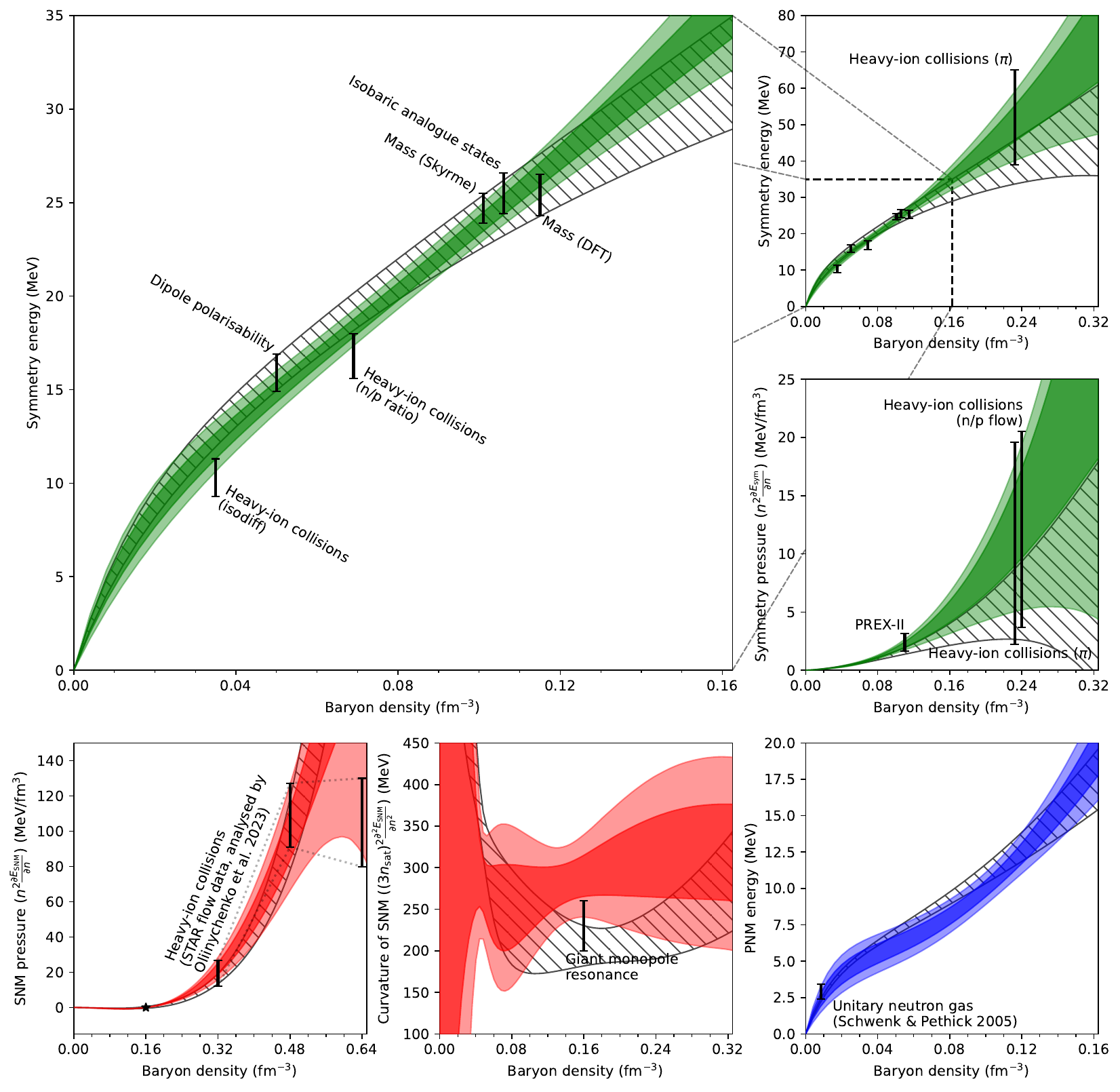}
    \caption{The `current experimental’ 1$\sigma$ constraints we use in this work, alongside the 1$\sigma$ and 2$\sigma$ bounds on the nuclear matter EOS distribution that results from applying them to our priors. All labels are the same as in Table 1 of \citet{tsang_determination_2024}, except for those drawn from other sources, for which references are given in the labels. For comparison, the 1$\sigma$ bounds for $\chi$EFT from \citet{lim_neutron_2018} are also shown as hatched regions. \textbf{Top left and top right:} the nuclear symmetry energy, with the top right plot showing a wider density range. \textbf{Middle right:} the contributions of the symmetry energy to the pressure (the symmetry pressure). \textbf{Bottom left:} the pressure of symmetric nuclear matter, with a marker indicating the fixed saturation density. \textbf{Bottom middle:} the curvature of the symmetric nuclear matter EOS. \textbf{Bottom right:} the EOS of pure neutron matter.}
    \label{fig:exp_constraints}
\end{figure*}

Figs.~\ref{fig:corner_JXEFT}, \ref{fig:corner_CXEFT} and \ref{fig:corner_exp} are corner plots showing the distributions of the isoscalar and isovector parameters resulting from the theoretical and experimental constraints we have used. Fig.~\ref{fig:exp_constraints} meanwhile shows where each of the experimental constraint we have used lies along the nuclear matter EOS, and the distribution for that EOS resulting from applying them to our priors.

\bibliography{paper6}